\documentclass[aps,twocolumn,pra]{revtex4}
\usepackage{bm}
\usepackage{graphicx}
\usepackage{epstopdf}
 
\RequirePackage[normalem]{ulem} 
\RequirePackage{color}\definecolor{RED}{rgb}{1,0,0}\definecolor{BLUE}{rgb}{0,0,1} 

\begin{document}

\title{Statistical physical theory of mode-locking laser generation with
a frequency comb}
\author{F. Antenucci$^{1,2}$, M. Ib\'a\~nez Berganza$^{3}$, L. Leuzzi$^{1,2}$} 

\affiliation{ $^1$NANOTEC-CNR, Institute of Nanotechnology, Soft and
  Living Matter Laboratory, Rome, Piazzale Aldo Moro 5, I-00185, Roma,
  Italy\\ $^2$ Dipartimento di Fisica, Universit\`a di Roma
  ``Sapienza,''Piazzale Aldo Moro 5, I-00185, Roma, Italy\\ $^3$INFN,
  Gruppo Collegato di Parma, via G.P. Usberti, 7/A - 43124, Parma,
  Italy}

\begin{abstract}
A study of the Mode-locking lasing pulse formation in closed cavities
is presented within a statistical mechanical framework where the onset
of laser coincides with a thermodynamic phase transition driven by the
optical power pumped into the system.  Electromagnetic modes are
represented by classical degrees of freedom of a Hamiltonian model at
equilibrium in an effective ensemble corresponding to the stationary
laser regime.  By means of optimized Monte Carlo numerical
simulations, the system properties are analyzed varying mode
interaction dilution, gain profile and number of modes.  Novel
properties of the resulting mode-locking laser phase are presented,
not observable by previous approaches based on mean-field
approximations.  For strong dilution of the nonlinear interaction
network, power condensation occurs as the whole optical intensity is
taken by a few electromagnetic modes, whose number does not depend on
the size of the system.  For all reported cases laser thresholds,
intensity spectra, phase waves and ultra-fast electromagnetic pulses
are computed.
\end{abstract}

\maketitle

\section{Introduction}

In multimode lasers with many cavity
modes, nonlinear interactions originate among modes.  One notable
mechanism inducing interaction is saturable absorption, i.e., the
progressive depletion of low power tails of the light pulse traveling
through the cavity at each roundtrip. This causes the consequent
amplification of very short pulses composed by modes with {\em locked
  phases}, a phenomenon called {\em mode-locking}
\cite{Haus00,HausBook}.  Mode-locking (ML) derives from the nonlinear
synchronization constraint on the oscillations of interacting
modes. Given any quadruplet of modes $\{k_1,k_2,k_3,k_4\}$ this is
expressed by the frequency matching condition (FMC):
\begin{equation}
\left| \nu_{k_1}-\nu_{k_2}+\nu_{k_3}-\nu_{k_4} \right| \leq  \gamma \ .
 \label{FMC}
 \end{equation}
 where $\gamma$ is the single mode line-width.  Phase-locking occurs
 at the ML lasing threshold and corresponds to some long-range order in
 the set of modes in the cavity.

We adopt a statistical mechanical approach to describe the optical
properties of stimulated light emission from cavities with a large
number of modes.  In this approach the generation of a multimode ML
lasing regime from a fluorescent continuous wave (CW) regime as the
optical power in the cavity is increased can be characterized as a
thermodynamic phase transition between a disordered phase and a phase
with long-range order .  The
stationary laser system can be treated as a thermodynamic system at
equilibrium in a thermal bath whose effective temperature is
proportional to the inverse squared power pumped into the cavity
\cite{Antenucci14,Marruzzo14}.  Since
the first attempt by Gordon and Fischer in the early 00's \cite{Gordon02}, this
approach has been performed in a mean-field fully connected
approximation corresponding in the optical language to the so-called
{\em narrow-band approximation}, see also Refs.  
[\onlinecite{Angelani06,Leuzzi09,Conti11, Antenucci14}]. This consists in choosing mode
frequencies in a narrow band-width $\Delta \nu$ around the central
frequency of the cavity. So narrow that the frequency interspacing
$\delta\nu$ between resonant modes is less than the linewidth $\gamma$
of each mode. In this way Eq. (\ref{FMC}) is practically always
satisfied and, therefore, actually irrelevant in determining lasing
properties.

In the present work we introduce frequency dependent populations of
modes, considering  gain profiles $g(\nu)$ and the effect of
nontrivial frequency matching on the mode couplings. This analysis requires to
go beyond the limits of validity of mean-field theory and it is
carried out by means of optimized Monte Carlo (MC) simulations running
on GPU's.  An exhaustive numerical analysis accounting for the
fluctuations induced by these new ingredients reveals that, depending
on the optical system properties, on the cavity topology, and on the
relative gain-to-nonlinearity strength, different thermodynamic-like
phases occur.  Such regimes range from a ferromagnetic-like one, where
all mode phases are aligned, to a {\em phase-wave} one, where phases
of modes at nearby frequencies are strongly correlated, though not
equal to each other.  The ferromagnetic behavior occurs in the low
finesse limit of the narrowband approximation.  Non-trivial phase
locking occurs, instead, at high finesse.  In the latter case we show how,
distributing the frequencies according to an optical frequency comb
\cite{Udem02, Baltuska03,Schliesser06}, intensity spectra and pulse
phase delay observed in ultra-short pulses are reproduced
\cite{Brabec00}.

As it will be taken up in the following, previous studies based on mean-field theory
 are exact only in the narrow band-width case. 
  In this paper we go beyond the mean-field approximation, accounting also for situations 
  in which different modes exhibit different frequencies

 Our study introduces two essential and new ingredients.  The first
 one is the FMC, yielding mode
 interaction networks that are no longer described by mean-field
 theory, in which non-trivial multimode emission spectra and mode
 phase correlations above threshold occur.  The second ingredient is a
 {\em random} dilution of the interacting network, modeling possible
 topological disorder in arbitrary cavity structures, as, e.g., multi
 cavity channels not exactly equal to each other.  We will show that,  as
 far as it is not too strong, the latter kind of dilution does not
 alter at all the laser transition properties. Below a certain dilution point,
 however, in the lasing phase the whole optical power condenses
 into a small set of connected modes, scaling independently of the number of
 modes.

\section{The Model }
Expanding the electromagnetic field in 
 the complete base of  $N$ normal modes $\{\bm E_n(\bm r), \nu_n\}$\cite{LambBook}

\begin{equation}
\bm E(\bm r, t)=\sum_{n=1}^N a_n(t) e^{-2\pi\imath \nu_n t} \bm E_n(\bm
r) + \mbox{c.c.}
\label{eq:Ert}
\end{equation}
the equilibrium dynamics of the time-dependent complex amplitudes
$a_n(t)$ is given by the Hamiltonian
\cite{Gordon02}
 \begin{equation}
{\cal H}=-\sum_{k=1}^N g_k |a_k|^2- J \sum
_{\{k_1,k_2,k_3,k_4\}}^{{\rm ML}}
a_{k_1}a_{k_2}^*a_{k_3}a_{k_4}^* 
\label{H}
\end{equation}
where $g_k$ and $J$ are chosen as real numbers, neglecting
dispersion and Kerr-lens effect.  The physical meaning of the
coefficients comes from the equivalence of the Hamiltonian dynamical
equation with the Haus master equation \cite{Haus00}: $g_k = g(\nu_k)$
is the net gain profile, $J$ is the self-amplitude modulation (SAM)
coefficient. The ML sum runs over a subset of quadruplets such that
for each element $(k_1,k_2,k_3,k_4)$ the FMC holds. The latter
implies that in the non-linear term of Eq. (\ref{H}) three
non-equivalent orderings of quadruplets contribute to the sum, each
one consisting of eight equivalent index permutations \footnote{The
  three orderings inequivalent with respect to the FMC are
  $\{k_1,k_2,k_3,k_4\}$, $\{k_1,k_3,k_2,k_4\}$ and
  $\{k_1,k_4,k_2,k_3\}$. Given a quadruplet $\{A,B,C,D\}$ the
  equivalent permutations are (i): $A \leftrightarrow C$, (ii): $B
  \leftrightarrow D$, (iii): $A \leftrightarrow B ~\&~ C
  \leftrightarrow D$ and their combinations.}.  The Hamiltonian is symmetrized with respect to these orderings.  The
coupling strength in Eq. (\ref{H}) is taken as $J=N/N_q$, where
$N_q$ is the number of quadruplets, making the Hamiltonian 
extensive.

The total optical energy stored in the system is ${\cal E}=N\epsilon =
\sum_{k=1}^N|a_k|^2$ and it is kept constant in the dynamics by
external power pumping.\ Eq. (\ref{H}) is a direct generalization of
the Hamiltonian studied in Ref.  \cite{Gordon02} and can be seen as
the ordered limit of the random laser theory analyzed in
Refs. \cite{Leuzzi09,Conti11, Antenucci14}.  From the point of view of
statistical mechanics the driven optical system composed by the
cavity, the amplifying medium {\em and} the optical power pumped into
the system can be described by Eq. (\ref{H}), considering it as the
Hamiltonian of a system at equilibrium with an effective thermal bath.
The role of the inverse temperature is played by the pumping rate
squared: ${\cal P}^2=\beta J\epsilon^2 $. Here $\beta=1/k_b
T$ is the inverse {\em heat bath} temperature, regulating
spontaneous emission. It is usually represented as white noise
in a Langevin dynamics \cite{Gordon02, Gordon03,
  Gat04,Angelani06, Leuzzi09,Conti11, Antenucci14, Marruzzo14}.

\section{Mode interaction network} 

Thermodynamic phases
are determined by the
interaction network, as well.  In
the following we will undergo the analysis of networks with a varying
degree of dilution. This will be expressed as number of quadruplets
$N_q$ vs. number of modes $N$.  We will discuss data for $N_q={\cal O}(N^t)$,
$t=1,2,3,4$.

Two essentially different types of topologies will be investigated, depending
on the frequency bandwidth being narrow or finite.  Both topologies
can be further diluted upon homogeneously randomly removing
quadruplets.  The ``Narrow Band-width Topology" (NBT) is {\em low finesse},
i.e., $\delta \nu \ll \gamma$, and the role of frequencies is
irrelevant. The fully connected instance, consisting in
$N_q=N(N-1)(N-2)(N-3)/8$ interacting quadruplets, corresponds to a
closed Fabry-Perot-like cavity where all longitudinal modes are
localized in the same spatial region.  Possible random diluted NBT's
correspond to more complicated geometries, including multi-channels
set-ups.  For finite bandwidth, instead, we will work in the {\em
  high-finesse} limit, $\delta \nu \gg \gamma$, with sets of
equispaced frequencies
\cite{Bellini00,Diddams00,Udem02,Baltuska03,Schliesser06}. We will
term this a ``Frequency Comb Topology" (FCT). In this case the list of
quadruplets is extracted from those nontrivially satisfying
Eq. \ref{FMC}: modes are not all equivalent to each other and
mean-field theory does not hold.

\section{Numerical Simulations and Data Analysis}

 We performed
extensive Monte Carlo simulations of equilibrium dynamics by means of
the exchange MC \cite{Hukushima96} algorithm and the synchronous,
fully parallel MC \cite{Peretto,Mahmoudi,Metz2008,Metz2009}. The
latter, indeed, remarkably turns out to reproduce a reliable dynamics
in the present model \cite{Antenucciprep14}.  In the NBT, system sizes
from $N=25$ to $500$ have been simulated for random dilutions of
$N_q={\cal O}(N^t)$, $t=2,3,4$. \footnote{We also tested the
  network-to-network fluctuations over different numbers of network
  realizations finding that the fluctuations $[\overline{O^2} -
    \overline{O}^2]^{1/2}$ of any obserables $O$ over the distribution
  of topologies are one order of magnitude lower than thermal
  fluctuations, $[\langle O^2\rangle - \langle O\rangle^2]^{1/2}$,
  already in the worst case of small size $N=100$.}.  For the FCT, we
simulated systems of size $N=100-1000$ with number of frequencies
$N_f=N$ in each case, and $N_q={\cal O}(N^2)$ and ${\cal O}(N^3)$ upon
applying the FMC filter.

The gain $g(\nu_n)$ is taken as Gaussian with varying mean square displacement. 
 We checked thermal
equilibration, i.e., the onset of the pumped stationary regime, by
looking at the energy relaxation and at the symmetry of the
distribution of complex amplitude values deep in the lasing phase.
In the following we present our results about (I) laser
thresholds identification, (II) intensity spectra, and (III)
phase waves, electromagnetic pulses and their correlations.

\subsection*{(I) Laser threshold} 
\begin{figure}[t!]
 \includegraphics[width=1.\columnwidth]{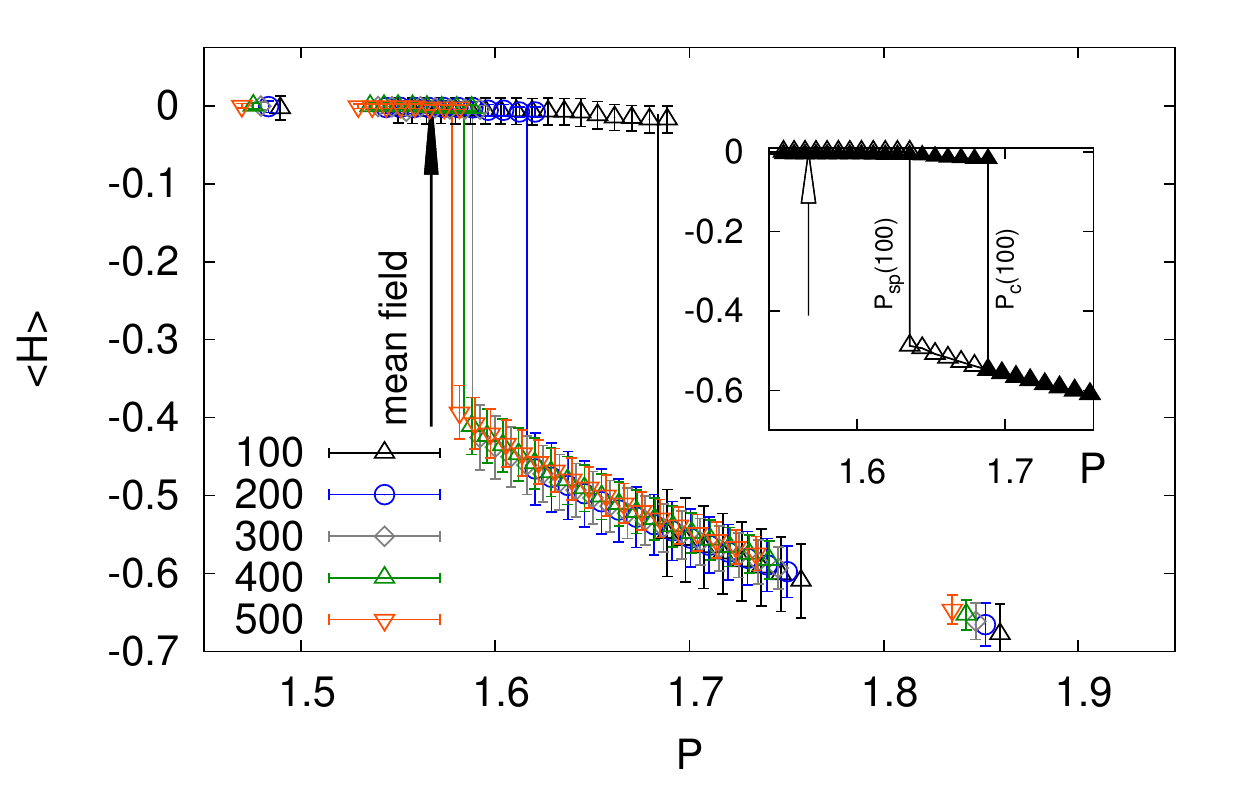}
 \caption{(Color online) Energy vs. $\cal {P}$ (in arbitrary units) in
   the Frequency Comb case with $N_q \propto N^2$. The arrow marks the
   analytic critical point in the thermodynamic limit of the
   NBT. Inset: for $N=N_f=100$ modes the spinodal line ${\cal P}_{\rm
     sp}$ is shown next to the threshold critical line.  }
\label{fig:energy}
\end{figure}

\begin{figure}[h!]
 \includegraphics[width=1.\columnwidth]{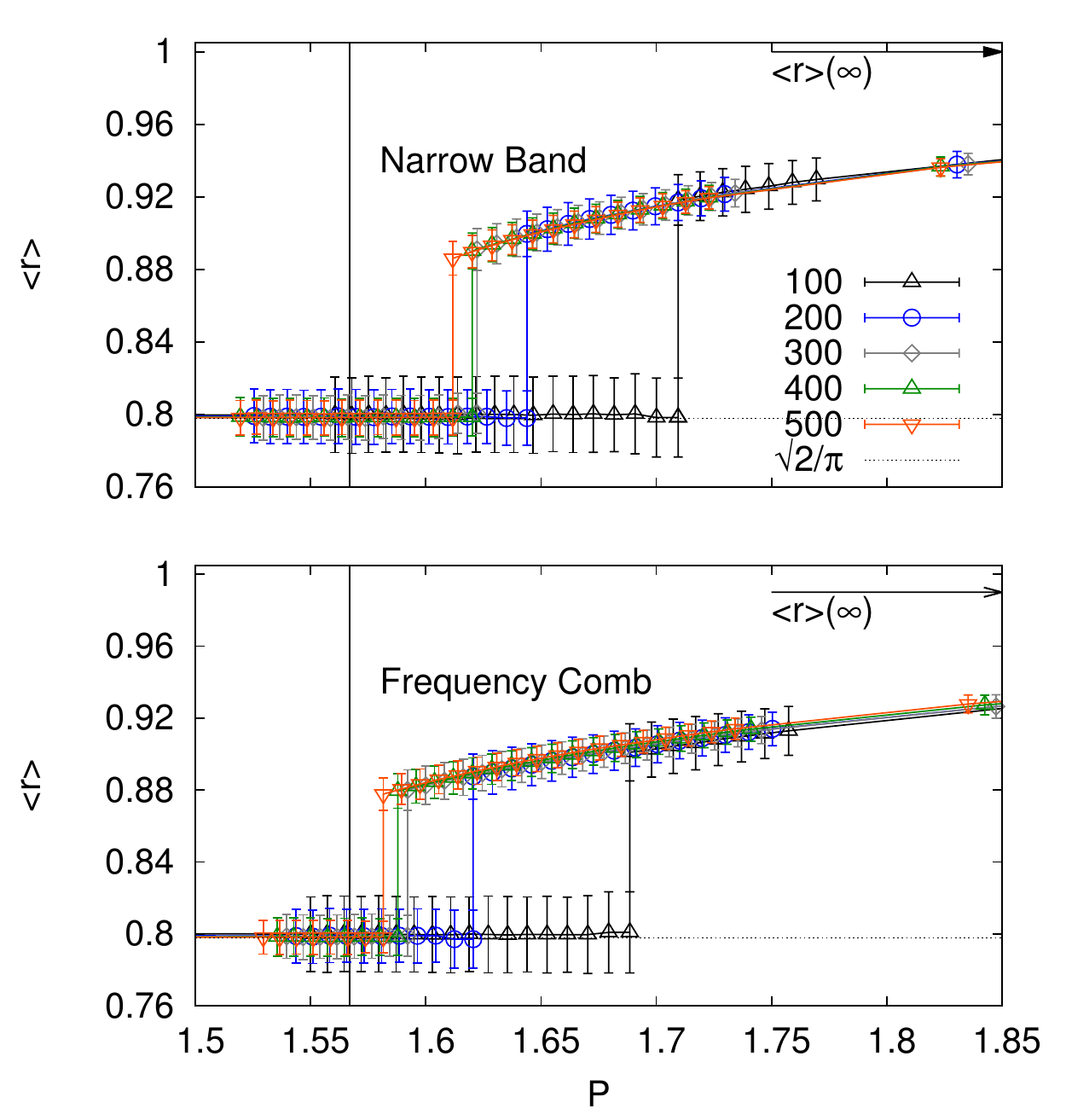}
\caption{ (Color online) Average mode magnitude $r=\langle |a|
  \rangle/\sqrt{\epsilon}$ vs. optical power ${\cal P}$ (in a. u.) for different
  sizes in the NBT (top panel) and in the FCT (bottom panel).}
\label{fig:radii}
\end{figure}

The estimate of the laser threshold is
obtained from the finite size scaling (FSS) analysis of the behavior
of the energy vs. pumping rate, as shown in Fig.  \ref{fig:energy} for
the FCT (for the NBT the energy behavior is the same). For low ${\cal
  P}$ the system is in an incoherent continuous wave regime with
uncorrelated phases and zero energy per mode.  As ${\cal P}$ increases
a phase transition occurs indicated by a discontinuity in the energy.
For the NBT the $N\to \infty$ critical point is analytically known
\cite{Antenucci14} and pointed out to as an arrow in
Fig. \ref{fig:energy}.  For $N_q={\cal O}(N^t)$, $t=2,3,4$ the FSS of
the discontinuity point is compatible with the fully connected
analytical limit, as reported in Tab. \ref{tab:pc}.  The critical
thresholds for the FCT case, estimate by FSS for $N_q={\cal O}(N^2)$
and for $N_q={\cal O}(N^3)$ are reported in Tab. \ref{tab:pc}.
The CW/ML laser phase transition is first order: in the inset of Fig. \ref{fig:energy} both the spinodal and the
critical points are displayed, e.g.,  for $N=100$ in a FCT.  Spinodal points
occur both in NBT and in FCT.
In Fig. \ref{fig:radii} the average mode magnitudes $r\equiv \langle
|a|\rangle/\sqrt{\epsilon}$ are plotted.  This is $\sqrt{2/\pi}$ for
randomly independently oscillating amplitudes and it discontinuously
increases at the ML lasing threshold indicating  intensity
mode-locking. For ${\cal P}\to\infty$, $r$ tends to $1$ in the NBT and to
$0.990(1)$ in the FCT case.

\begin{table}[b!]
\begin{tabular}{|c|cccc|cc|}
\hline
$\nu$ band & &Narrow&Band&  &Frequency &Comb \\
\hline
$O(N_q) $ &  $N^2$ & $N^3$ & $N^4$ & Exact & $N^2$ & $N^3$\\
\hline
${\cal P}_c$&$1.56(3)$&$1.59(9)$& $1.6(3)$ &$1.56697$& $1.558(8)$ & $1.57(1)$ \\
\hline
\end{tabular}
\caption{Critical point for $N\to\infty$ in various  dilutions.}
\label{tab:pc}
\end{table}

 \subsubsection*{Power condensation}
  As the dilution is strong, i.e., 
  $N_q={\cal O}(N)$, each mode interacts in a ${\cal O}(1)$ number of
 quadruplets.  Above threshold the whole power ${\cal{
     E}}$  turns out to be taken by a small number of connected modes
      and the probability to find
 a configuration with energy equipartition is negligible in the
 thermodynamic limit.
In the mean field
 approximation one can prove that in order to display power condensation it must be 
   $N_q < {\cal O}(N^2)$ \cite{Antenucciprep14} as
 confirmed by numerical simulations.  In the following we focus on
 more connected networks.

\begin{figure}[t!]
\includegraphics[width=\columnwidth]{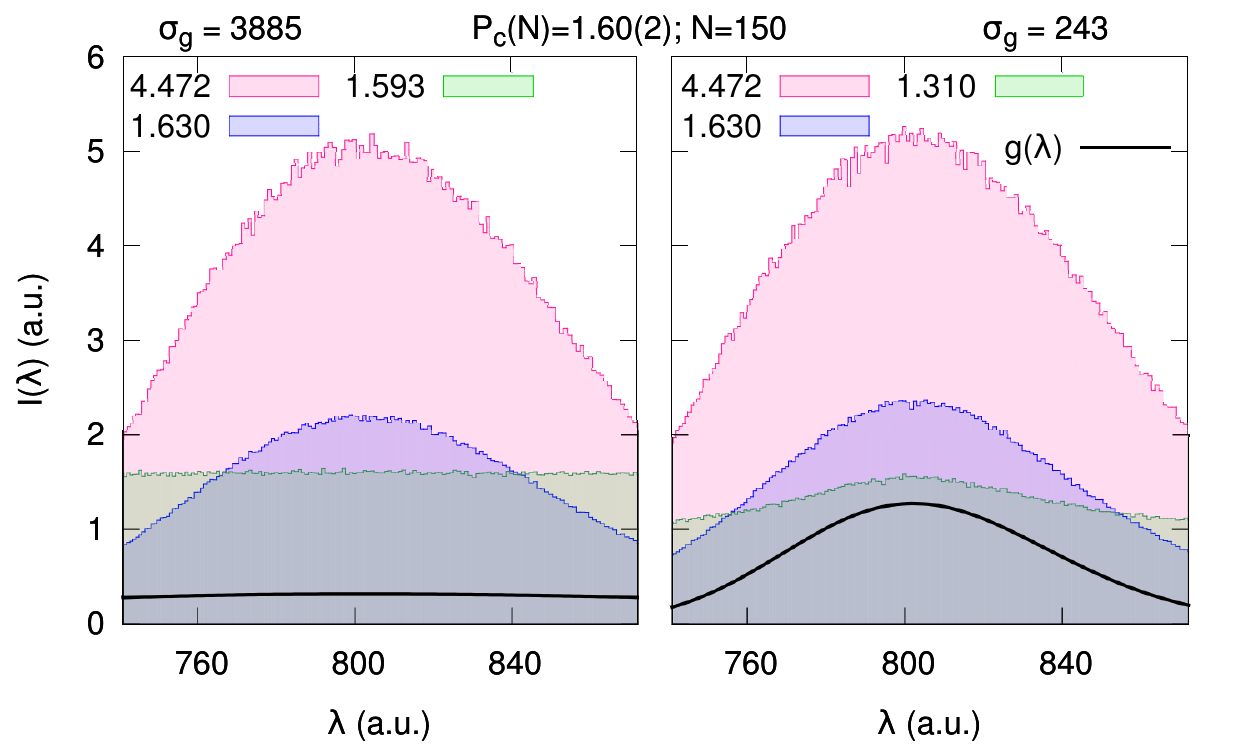}
\caption{(Color online) Intensity spectra for a FCT system
   of $N_f=N=150$. $N_q={\cal O}(N^2)$ for increasing ${\cal P}$ from bottom to top. Left: gain $g(\lambda)$ with
  larger variance, $\sigma_\lambda=3885$.  At ${\cal P}>{\cal P}_c$ the spectrum
  starts narrowing because of the nonlinear mode-coupling.  Right:
  $g(\lambda)$ with smaller variance, $\sigma_\lambda=243$. Spectra
  follow the peaked gain profile already in the CW regime. At ${\cal P}_c$
 mode-locking sets in, enhancing the sharpening.}
\label{fig:spectra}
\end{figure}

\begin{figure*}[t!]
\includegraphics[width=.95\textwidth]{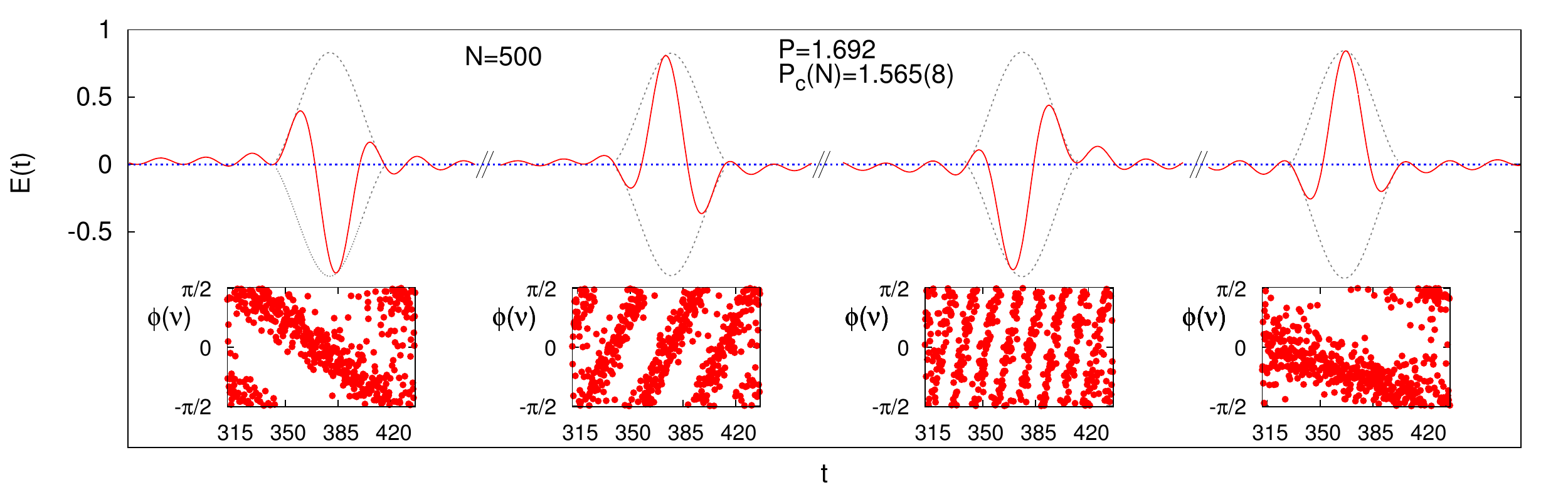}
\caption{
Electromagnetic field $E(t)$ at different emissions in the
  system dynamics with a uniform comb distribution for mode
  frequencies.  $N=N_f=500$.  Bottom insets: phase-locked linear
  behavior $\phi(\nu)$ corresponding to each pulse.  The phase
  shift in the peak of $E(t)$ with respect to the maximum of
 the envelope corresponds to the slope of
  $\phi(\nu)$. Time is in arbitrary units.}
\label{fig:pulse}
\end{figure*}

\subsection*{(II) Intensity spectra}
 In Fig. \ref{fig:spectra} we show
two instances of the spectra $I(\lambda_j)=\langle |a_j|^2\rangle$
vs. $\lambda_j=c/\nu_j$ in FCT systems with Gaussian gain profiles
with different variances.
 In the left panel the mean square displacement of the gain profile in
 the wavelength dominion is large ($\sigma_g=3885$) in
 comparison to the spectral free range, whereas in the right panel
 it is of the same order of magnitude ($\sigma_g=243$). In the
 first case, below ${\cal P}$ the CW spectrum is flat and suddenly
 sharpens at the ML threshold ${\cal P}_c$. To underline this, spectra are shown
 right below and  above ${\cal P}_c(N=150)=1.597(15)$ in  Fig. \ref{fig:spectra}.  In the small $\sigma_g$ case
 the spectra appears already narrower in the CW regime, following
 $g(\lambda)$, as displayed in the right panel of
 Fig. \ref{fig:spectra} for the lowest simulated pumping rate. At
 ${\cal P}_c$ though, their narrowing qualitatively changes and
 becomes progressively independent of $g(\lambda)$ as ${\cal P}$
 increases, eventually taking the same spectral shape of the previous
 case.

We show in Fig. \ref{fig:spectra}  the cumulative detections
of very many pulses, as in data acquisition from ultra-fast ML lasers. In
the MC dynamics used in simulation, though, each MC step
 corresponds to a pulse generation. Within our approach it is,
then, possible to look at the  dynamics at much shorter time
intervals, where the mode amplitude and intensity profile in  $\lambda$ fluctuates from pulse to pulse. 
This is connected to changes in the
spectral phase delay of the electromagnetic pulse. Different spectral
dynamics are reported in Video $1$ of Sup. Mat. \cite{SuMa} (see
details in Sec. \ref{SupMat}).

\subsection*{(III)  Electromagnetic pulses and phase delay}

In terms of slow complex amplitudes, cf. Eq (\ref{eq:Ert}), $a_n(\tau) = A_n(\tau)
e^{\imath \phi_n(\tau)}$, $A_n=|a_n|$, the electromagnetic pulse is
\begin{equation}
\nonumber E(t|\tau) =\sum_{n=1}^N A_n(\tau) e^{\imath [2\pi\nu_n t +
  \phi_n(\tau)]}
\end{equation} 
\\
\indent
The time $\tau \gg t$ operatively labels a single MC step in our
simulations, i.e., the interval between two pulses.  In Fig. \ref{fig:pulse} we show $E(t|\tau)$
at four different times $\tau$ in the dynamics.  In the NBT, in the ML
regime all modes acquire same modulus and phase.  In
a FCT, instead, at ${\cal P}_c$ a non-trivial phase-locking
occurs, such that the mode phases exhibit a linear dependence on the
mode frequencies: $\phi_n \simeq \phi_0+ \phi' \nu_n$, as shown in the
bottom insets of Fig. \ref{fig:pulse}. The pulse is, thus, unchirped
\cite{Haus00}.  The spectral {\em phase delay}, or group delay,
$\phi'= d\phi(\nu)/d\nu |_{\nu=\nu_n}$ of the optical pulse does not
depend on the frequency of mode $n$. It changes, though, with time $\tau$,
from pulse to pulse.  Within our approach we thus find the typical
spectral phase frequency profile $\phi(\nu)$ at each given pulse and
its pulse-to-pulse dynamics, cf. Video $2$ in Sup. Mat. \cite{SuMa}
(see details in Sec. \ref{SupMat}).

\subsubsection*{Phase waves lifetime}

\begin{figure}[t!]
 \includegraphics[width=\columnwidth]{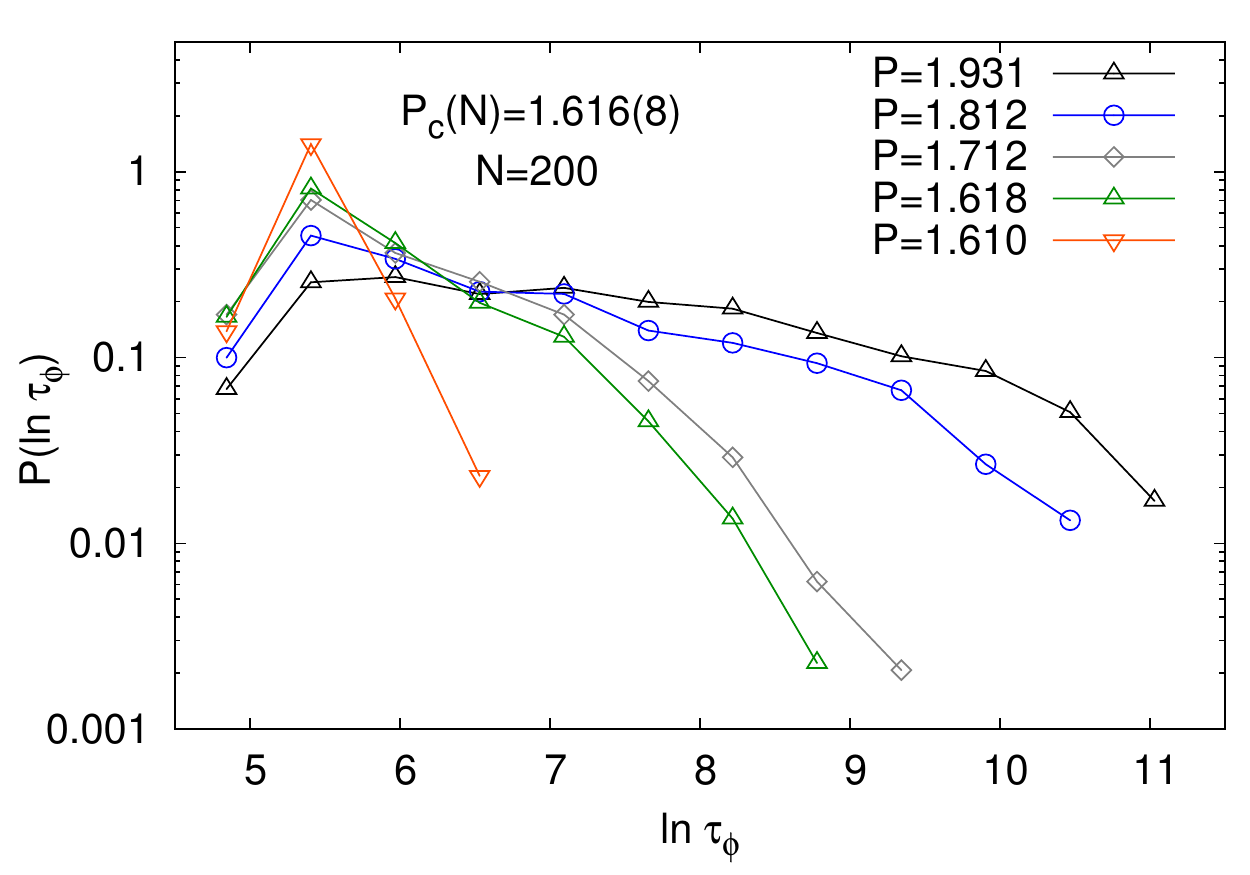}
\caption{(Color online) Distribution of the decay time of the equal
  time phase correlation function for a system of $N=N_f=200$ modes
  across the threshold ${\cal P}_c(N=200)=1.616(8)$. Time $\tau$ in is
  Monte Carlo steps.}
\label{fig:TAU_PHI}
\end{figure}

 Let us define the time average over
an equilibrated set of data ($\tau \geq \tau_{\rm therm}$) on a
time window ${\cal T}$: $\langle \ldots \rangle_{\cal T}
\equiv \sum_{\tau=0}^{\cal T}(\ldots)/{\cal T}$.  In the FCT, after a
time ${\cal T}>\tau_\phi$, the average global phase correlation function $ {\cal C}_\phi ({\cal T})$,
defined as
\begin{eqnarray}
{\cal C}_\phi({\cal T}) &\equiv& \frac{1}{N_f}\sum_{\delta\nu} 
\left| C_{\delta\nu}({\cal T})
\right| 
\label{phicorr}
\\ 
C_{\delta\nu}({\cal T})&=&\frac{1}{N_f}\sum_{\nu}\langle
\cos \left( \phi_\nu - \phi_{\nu+\delta\nu} \right) \rangle_{{\cal T}}
\label{nucorr}
\end{eqnarray}
 is observed to decay to zero. 
This is at difference with the NBT
 where, in the high power regime, $ C_\phi ({\cal T})$ is finite also
 for ${\cal T}\to\infty$.   For the FCT, the distribution of correlation times
 $\tau_\phi$ as the optical power varies across the lasing
 threshold is sharply peaked around its logarithmic average
 ${\overline{\ln \tau_\phi}}$ below threshold, cf.
 Fig. \ref{fig:TAU_PHI}.  For increasing ${\cal P}>{\cal P}_c$ the
 distribution tends to a flat curve.

 \begin{figure}[t!]
\includegraphics[width=0.9\columnwidth]{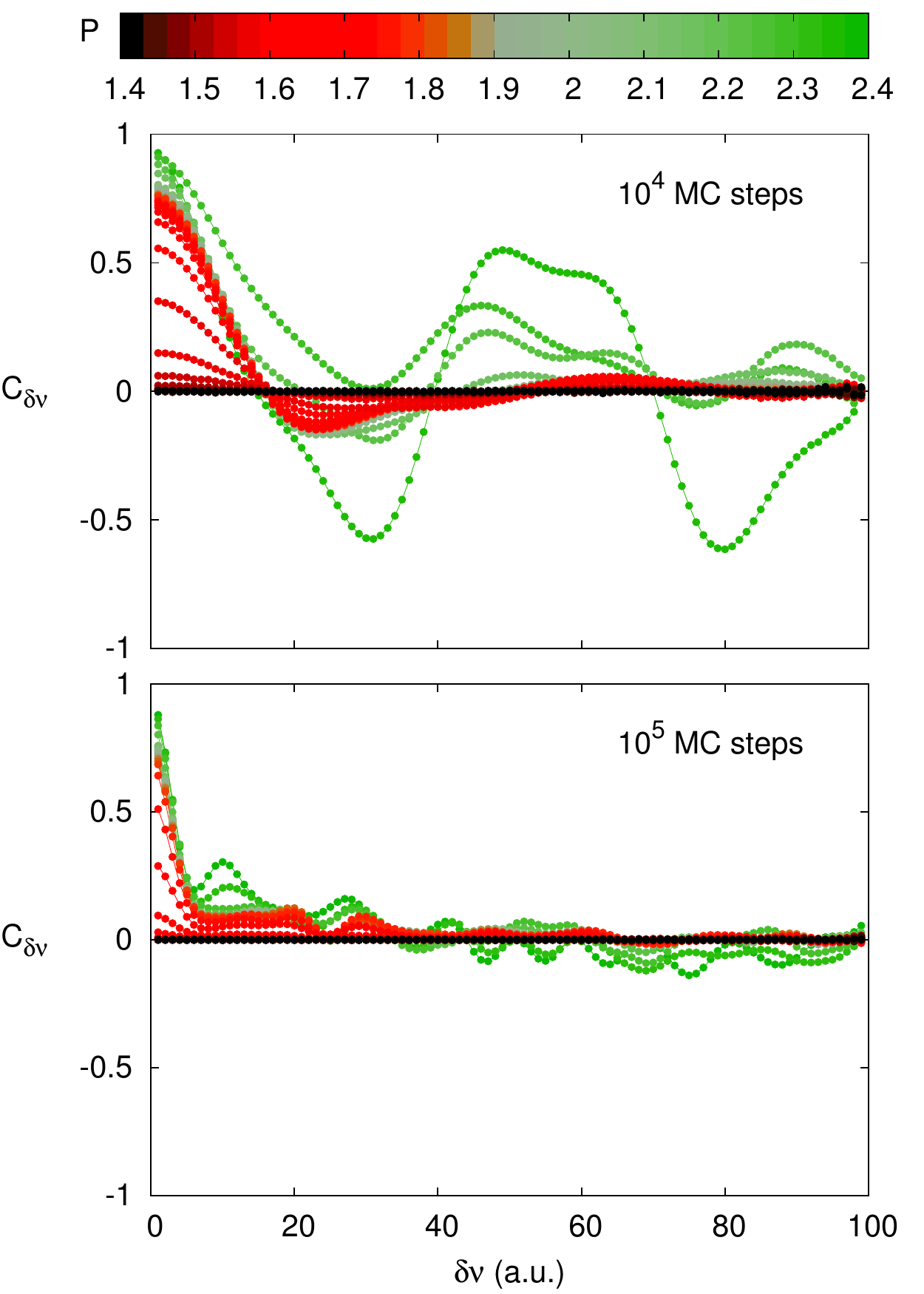}
\caption{(Color online) Phase-Phase correlation $C_{\delta\nu}({\cal
    T})$ as function of frequency difference $\delta\nu=\nu'-\nu$ and
  optical power ${\cal P}$ for ${\cal T}=10^4$ (top panel) and ${\cal
    T} = 10^5$ (bottom panel) in a FCT cavity with $N=100$ and $N_q =
  \mathcal{O}(N^2)$. The threshold power is ${\cal P}_c(100)=1.62(2)$
  and the range of power of the displayed correlations is reported on
  the top palette from black (black online) for low power to light
  grey (light green online) for high power. }
\label{fig:corr_phi_nu}
\end{figure}

\subsubsection*{Vanishing two-mode correlators}

 A related phenomenon is that
the average over ${\cal T}> \tau_\phi$ of two-mode phase correlations
$C_{\delta\nu}({\cal T})$, cf. Eq. (\ref{nucorr}), vanishes, as shown in
Fig.  \ref{fig:corr_phi_nu}, implying a zero ensemble average.  This
occurs though modes with frequencies $\nu$, $\nu'=\nu+\delta\nu$ {\em are} 
correlated at each time $\tau$ , cf.  bottom insets of
Fig. \ref{fig:pulse}, and $C_{\delta\nu}({\cal T})\neq 0$ when ${\cal
  T}\lesssim\tau_\phi$.
In Fig. \ref{fig:corr_phi_nu} we show  $C_{\delta\nu}({\cal T})$ as
function of $\delta\nu=\nu'-\nu$ for ${\cal T}=10^4$ and ${\cal
  T}=10^5$.  In the top panel, at shorter time window ${\cal T}=10^4$,
one clearly observes that $C_{\delta\nu}$ is completely uncorrelated
independently of $\delta\nu$ for ${\cal P}<{\cal P}_c$. As the pumping
increases above the threshold, $C_{\delta \nu}$ displays a non-trivial
behavior as a function of $\delta\nu$. Above threshold, thus, the
global phase correlation function ${\cal C}_\phi$, cf.
Eq. (\ref{phicorr}), becomes larger the higher the pumping.  In other
words, the correlation time $\tau_\phi$ grows with ${\cal P}$ and
overcomes ${\cal T}$: $\tau_\phi({\cal P}>{\cal P}_c)> 10^4$.  In the
bottom panel of Fig. \ref{fig:corr_phi_nu} we consider, instead, a
time window ${\cal T}$ larger than the average correlation time
$\tau_\phi({\cal P})$ for most of the simulated pumping values ${\cal
  P}$, cf. Fig. \ref{fig:TAU_PHI}: it can be observed that
$C_{\delta\nu}(10^5)\simeq 0$ for practically almost all $\delta\nu$,
but the smallest ones for large power.
The oscillations displayed by $C_{\delta\nu}({\cal T})$ in Fig.
\ref{fig:corr_phi_nu} in the ML laser regime are due to the fact that
different phase delays are involved in the thermal average. Indeed,
cf. bottom insets of Fig. \ref{fig:pulse}, the slope of $\phi(\nu)$
changes with time $\tau$.

The origin of the vanishing of two-mode correlators is reminiscent of
symmetry conservation in gauge lattice theories \cite{Kogut79} and
will be discussed elsewhere \cite{Antenucciprep14}. We just mention
that the main difference in the lasing regime for the two topologies
is that in the NBT the global $U(1)$ symmetry is spontaneously broken,
whereas in the FCT it is conserved across the threshold.

\section{ Conclusions }

We present the first statistical
mechanical approach to the study of real-world ultrashort mode-locked
multimode lasers in closed optical cavities, including possible
degrees of topological disorder. 
In previous  approaches, statistical mechanical systems with distinct resonances 
have been studied in the mean-field approximation, see, e. g., Refs. 
\cite{Gordon03,Gat04, Rosen10}.
The key point is, though,  that  the mean field solution is exact
only in the narrow band-width limit. 
When describing inhomogeneous topologies, such as the Frequency Comb
Topology of equi-spaced well-refined resonances,
the only thing that the mean-field theory can account for is a shift
in the pumping threshold resulting from the dilution (in fact, just a modification of the 
coupling constant). The nature
of the predicted mode-locked regime remains, indeed, identical to the one predicted
assuming narrow band-width.
This limit  basically lies in the very definition of the mean-field method: 
since the fluctuations of the mode degrees of freedom are neglected
 a many-body problem is actually reduced to a one-body problem,
 in which all modes exhibit   a common average phase and
a common average intensity.
The inhomogeneity in frequency dependence of the interaction network in more realistic cases,
 in which modes with near-by frequencies
 have stronger coupling, is simply neglected by construction.
Because of the mean-field assumption, previous 
approaches have not, and could not have, 
accounted for the main
properties here reported: phase
waves, non-equipartition threshold and vanishing two-mode
correlators. 

Our approach, going beyond mean-field theory, with
Monte Carlo simulations of equilibrium dynamics, allows
to reproduce and study the onset of the lasing regime and the behavior
of emission spectra and laser pulses and relative group phase delays at any supplied power.  The
existence of metastable lasing regimes marked by spinodal points in
the energy behavior, cf. inset of Fig. \ref{fig:energy}, accounts for  the
onset of optical bistability \cite{Gibbs85,Baas04}. The
phenomenon of power condensation for extreme dilution of mode
interaction and the vanishing of the equal time two-mode phase
correlations for long times are properties that can be experimentally tested.  Furthermore, this kind of
approach opens the way to further analyze the carrier-envelop offset
phase behavior, and the tolerance to disorder in the coupling SAM
coefficient.  The latter analysis is useful, e.g., for stabilized
micro resonator in chip-based devices \cite{Lee12,Saha13} in which
technical precision undergoes $\mu m$ size constraints and controlling
material damage is a true challenge. Eventually, including open cavity
terms \cite{Viviescas03,Hackenbroich03} and strong disorder in the
nonlinear coupling \cite{Angelani06, Leuzzi09,Conti11,Antenucci14},
our approach can be applied to the study of random lasers
\cite{Lawandy94,Cao98,Cao05,Wiersma08, Ghofraniha15}.

\section*{Acknowledgements} The authors would like to thank Claudio Conti, 
Andrea Crisanti and Giorgio Parisi
 for stimulating discussions.  The research leading to these results has
received funding from the Italian Ministry of Education, University
and Research under the Basic Research Investigation Fund (FIRB/2008)
program/CINECA grant code RBFR08M3P4 and under the PRIN2010 program,
grant code 2010HXAW77-008 and from the People Programme (Marie Curie
Actions) of the European Union's Seventh Framework Programme
FP7/2007-2013/ under REA grant agreement n¡ 290038, NETADIS project.

\section{Supplemental Material}
\label{SupMat}

\subsection*{Spectra dynamics [Video 1]}

In the file {\tt {Video1\_spectral\_dynamics.mpg}} a video is shown for the
dynamics of spectra $I(\lambda;\tau)=|a(\lambda;\tau)|^2$ for a FCT system with
$N=N_f=500$ modes and frequencies. The gain wavelength profile
$g(\lambda)$ is taken uniform, i.e. $\sigma_g=\infty$. The finite size
threshold for this specific system is ${\cal P}_c=1.578(8)$.  Dynamic
sequences at three pumping rate are reported: in the incoherent CW
regime (${\cal P}=1.308$), slightly above the lasing threshold (${\cal
  P}=1.865$) and for high pumping (${\cal P}=2.94$). Each MC step
corresponds to the interval between two pulsed emissions in the
mode-locked lasing regime. In the video each single frame is averaged
over 10 subsequent Monte Carlo steps and frames are shown at intervals
of 100 Monte Carlo steps.

\subsection*{Lasing pulse and phase delay dynamics [Video 2]}

In the file {\tt{ Video2\_phase\_delay\_dynamycs.mpg}} a video is shown for
the stationary dynamics of the relationship $\phi(\nu)$ of lasing
pulses in a  simulation of a FCT system of $N=N_f=500$
modes with an initial random dilution of $N_q^{\text{tot}} = 37500000
= 0.3  N^3$ and a final number of interacting quadruplets 
$N_q^{\text{FMC}} = 49965$. The lasing system is at optical power ${\cal
  P}=1.651$, right above the lasing threshold ${\cal P}_c=1.578(8)$.
The initial distribution of the gain among frequencies is taken as
uniform. The interval between each frame is 100 Monte Carlo steps.  It
can be observed that at each time $\tau$, corresponding to a pulsed
emission, $\phi(\nu)$ is approximately linear, yielding a well defined
phase delay $\phi'$ independent from $\nu$. As the dynamics runs,
though, $\phi'(\tau)$ changes, progressively taking a broad interval
of values.

\end{document}